# Chemisorption of a molecular oxygen on the UN (001) surface: *ab initio* calculations

Yu.F. Zhukovskii, D. Bocharov[*], and E.A. Kotomin

Institute of Solid State Physics, University of Latvia, 8 Kengaraga Street,
LV-1063 Riga, Latvia

**Abstract:** The results of DFT GGA calculations on oxygen molecules adsorbed upon the (001) surface of uranium mononitride (UN) are presented and discussed. We demonstrate that $O_2$ molecules oriented parallel to the substrate can dissociate either (*i*) spontaneously when the molecular center lies above the surface hollow site or atop N ion, (*ii*) with the activation barrier when a molecule sits atop the surface U ion. This explains fast UN oxidation in air.

## 1. Introduction

The uranium mononitride (UN), which possesses a rock salt structure and metallic nature, is an advanced material for the non-oxide nuclear fuel considered as a promising candidate for the use in Generation-IV fast nuclear reactors [1]. UN reveals several advantages over a traditional $UO_2$–type fuel (*e.g.*, higher thermal conductivity and metal density). However, one of important problems with actinide nitrides is their effective oxidation in contact with oxygen which can affect nuclear fuel performance [2].

There was a series of *ab initio* density functional theory (DFT) calculations published in last 10 years on pure and defective $UO_2$ (*e.g.*, [3-10]). Similar calculations on the UN appeared only recently [11-16]. In our recent papers, we studied both the structure of a perfect UN(001) surface [17] and chemisorption of oxygen atoms upon it [18]. These DFT calculations were performed using the two quite different computer codes: *VASP 4.6* [19], with plane wave basis set (BS), and *CRYSTAL-06* [20], with the BS of localized atomic orbitals (LCAO approach). In both cases we have applied the non-local exchange-correlation functional by Perdew-Wang-91 (PW91), that is, the generalized gradient approximation (GGA) [21]. The results of these two different methods reveal good agreement [17,18] which supports their reliability. A strong *chemisorption* was observed for O atom interaction with the UN surface (~7 eV atop the surface U ion) which is typical for traditional metallic surfaces (*cf.* ~10 eV *per* adatom bound on the close-packed Al surfaces [22]). However, to shed more light on understanding the UN oxidation mechanism, we theoretically study in this paper the interaction of *molecular* oxygen with the same defectless UN(001) surface. The key questions are: whether the $O_2$ dissociation upon the surface is energetically possible, which adsorption sites are optimal for this, and whether it can occur spontaneously, without energy barrier. There are important issues for understanding the mechanism of the oxidation of uranium nitride in air.

## 2. Theoretical

We have employed the *VASP 4.6* code [19] with the scalar relativistic PAW pseudopotentials representing the core electrons of U ($6s^2 6p^6 6d^2 5f^2 7s^2$ valence shell), N ($2s^2 2p^3$) and O ($2s^2 2p^4$) atoms as well as the non-local PW91 exchange correlation functional [21]. The cut-off energy has been chosen 520 eV. We have applied the Monkhorst-Pack scheme [23] with 4×4×1 *k*-point mesh in the Brillouin zone (BZ). When modeling the UN(001) surface, we have used the same 3D symmetric slabs as previously [17,18] consisting of five non-polar layers, containing alternating U and N atoms, separated by large vacuum gaps along the *z* axis (~36 Å) and thus excluding the direct interaction of oxygen molecules from the neighboring slabs. The lattice constant (4.87 Å) optimized for the bulk has been used in all our slab calculations. Only the *ferromagnetic* ground state has been considered in this study as the energetically most preferable at low temperatures.

---

[*] Corresponding author: Tel.: +371 67187480, Fax.:+371 67132778
E-mail address: bocharov@latnet.lv (D. Bocharov).

For simulation of the chemisorption of oxygen molecule, we have used the 2×2 extended surface supercell (containing 20 U cations and 20 N anions), similarly to the previous study on chemisorption of an atomic oxygen [18]. The periodic adsorbate distribution corresponds to the molecular coverage of 0.25 ML (or atomic O coverage of 0.5 ML). To reduce computational efforts, we have used a symmetric two-sided arrangement of oxygen molecules. The binding energy $E_{bind}$ *per* oxygen atom in the adsorbed molecule $(O_2)_{ads}$ was calculated as:

$$E_{bind} = \frac{1}{4}\left(E^{UN} + 2E^{O_2} - E^{O_2/UN}\right), \qquad (1)$$

where $E^{O_2/UN}$ is the total energy of a fully relaxed $O_2$/UN(001) slab for several configurations of $(O_2)_{ads}$ upon the substrate (with a center of molecule atop the corresponding surface site as shown in Fig. 1), $E^{O_2}$ and $E^{UN}$ the total energies of an isolated oxygen molecule in the ground (triplet) state and of a pure relaxed slab, respectively. The factor 1/4 before brackets appears since the substrate is modeled by a slab containing the two equivalent surfaces with $(O_2)_{ads}$ positioned symmetrically relatively to both slab surfaces whereas each molecule before and after dissociation contains two oxygen atoms.

## 3. Main results

When modeling the molecular adsorption, we have analyzed different configurations of an $O_2$ molecule in the triplet state on the UN(001) substrate. *Vertical* orientations of the molecule atop the surface N or U ions have been found metastable with respect to molecule reorientation to the *horizontal* configuration, parallel to the surface. We have estimated both the binding energy of a molecule using Eq. (1) and the molecule dissociation energy (for some configurations), *i.e.*, the difference of the total energies of a slab with an $O_2$ molecule before and after dissociation, when the two O atoms in the triplet state sit atop the two nearest surface U ions (Table 1).

*3.1. Spontaneous dissociation of $O_2$ molecules*

We have found that a spontaneous, barrierless $O_2$ dissociation indeed takes place in the two cases: when the molecular center is atop either (*i*) a hollow site or (*ii*) surface N ion, with the molecular bond directed towards the two nearest surface U ions (the configurations 1 and 5 in Fig. 1, respectively). The relevant dissociation energies $E_{diss}$ are given in Table 1, along with other parameters characterizing the atomic relaxation and the Bader charge distribution. Geometry and charges for the configurations 1 and 5 after dissociation (Table 1) are in general similar to those obtained in our previous study [18] for UN(001) substrate covered by chemisorbed O atoms, *e.g.*, surface U atoms beneath the oxygen adatom after dissociation are shifted up in both configurations (Table 1). However, since concentration of $O_{ads}$ in this study is twice larger as compared to that for atomic oxygen [18], some quantitative differences in the results presented in Tables 1 of this paper and Ref. [18] are unavoidable. For example, the *repulsion* energy between the two adatoms after $O_2$ dissociation sitting atop the two nearest surface U ions (the configuration 1) is quite noticeable, ~0.7 eV.

We have also identified two other configurations of adsorbed oxygen molecules where the dissociation is energetically possible but with the energy barrier: (*i*) atop the hollow site when a molecular bond is oriented towards the nearest N ions (the configuration 2 in Fig. 1) and (*ii*) atop the surface U ion (for any molecular orientation, *e.g.*, the configurations 3 and 4 in Fig. 1). For the configuration 2, we have observed the orientation instability of the adsorbed molecule which easily rotates, *e.g.*, towards the surface U ion with further dissociation. The configurations 3 and 4 could be characterized as rather metastable $UO_2$ *quasi-molecules* due to a strong bonding between all three atoms (Fig. 2c) and since the corresponding U ion is noticeably shifted up from its initial positions on surface (Table 1). Meanwhile, the dissociation of $(O_2)_{ads}$ molecule in the configuration 3 is energetically possible but only after overcoming the activation energy barrier.

*3.2. Charge redistribution analysis*

Adsorption of an $O_2$ molecule (in the triplet state) is accompanied by the charge transfer of ~1 *e* (*per* molecule) from the substrate (Table 1). In Fig. 2 we analyze the difference electron charge redistributions for three configurations of horizontally oriented $(O_2)_{ads}$ upon the surface: (*a*) molecule adsorbed upon the hollow site (the configuration 1, Fig. 1), (*b*) molecule dissociated from this configuration with O adatoms located atop the nearest surface U ions, and (*c*) molecule adsorbed upon the surface U ion (the configuration 3). Spontaneous $O_2$ dissociation and thus a smooth transition from the charge distribution (a) to (b) can be explained by continuous areas of the electron density (Fig. 2a) parallel to the surface which may be considered as *dissociation channels*, analogously to the density plot for a molecular oxygen upon the Al substrate [22]. After dissociation each O adatom contains an extra charge of ~1 *e*, *i.e.* transforms into $O^-$ ion in the triplet state (Fig. 2b). In contrast, when considering the molecular configuration 3, these *dissociation channels* are transformed into *dissociation barriers* (Fig. 2c). Simultaneously, we observe considerably higher electron density, indicating a kind of $UO_2$ quasi-molecule with a strong bonding between the $O_2$ molecule and surface U atom beneath. Thus, difference between the electron density plots presented in Figs. 2a and 2c can explain different dissociation abilities of $O_2$ molecule in the configurations 1 and 3 (Fig. 1).

*3.3. Electronic densities of states (DOS)*

For the same adsorbate configurations considered above, we have constructed the total and projected densities of states (DOS) (Fig. 3). Molecular adsorption in these configurations leads to appearance of the specific *oxygen bands* as compared to those for oxygen adatoms upon UN surface [18] and O atom substituted for a host N ion in UN bulk [15]. For a molecular oxygen atop the hollow position (Fig. 3a), O *2p* peak is observed at -1 eV overlapping with the U *5f* and *6d* bands. After $O_2$ dissociation (Fig. 3b) this peak disappears being replaced by the broad two-peak band in the region of the N *2p* valence band (-2 to -5 eV), similarly to the DOS for oxygen adatoms on UN(001) substrate [18]. Some differences are also noticeable between the corresponding U *5f* and *6d* peaks in the spectral range above -1 eV (*cf.* Figs. 3a and 3b) which could be caused by both different arrangement of O and U atoms in these configurations and sensitivity of uranium states to the presence of oxygen, thus indicating once more a strong oxygen chemical bonding (chemisorption). When oxygen molecule is located atop the surface U ion (the configuration 3), the U *5f* and *6d* contributions in the energy range above -1 eV are diminished, simultaneously the O *2p* contribution grows, thus increasing an overlap between all three states and indicating $UO_2$ quasi-molecular bond formation. As compared to the adsorption of oxygen molecule upon the hollow site (Fig. 3a), we again observe a higher O *2p* peak (at -1.5 eV) and an additional lower peak of the same O *2p* (at -5.5 eV) which noticeably overlaps with the U *5f* and *6d* subpeaks (Fig. 3c). Some analog of the latter pattern was observed earlier for the projected DOS of O atom substituted for N in UN bulk [15]. In all three DOS (Fig. 3), a broad band corresponding to the N *2p* projected states does not change drastically which means a weak effect of N ions on the $O_2$ molecule adsorption on the UN(001) surface.

## 4. Conclusions

Summing up, the results of our *ab initio* calculations clearly demonstrate a real possibility for spontaneous dissociation of the adsorbed oxygen molecules upon the perfect UN(001) surface, analogously to the $O_2$ dissociation on "traditional" metallic surfaces. This is the first important step in understanding the initial stage of the UN oxidation mechanism.

**Acknowledgements**

This study was partly supported by the European Community FP7 project F-Bridge. D.B. gratefully acknowledges also the support from the European Social Fund (ESF). The authors kindly thank R. Caciuffo, R.A. Evarestov, D. Gryaznov, E. Heifets, Yu. Mastrikov, Hj. Matzke, and P. Van Uffelen for fruitful discussions.

**Figures**

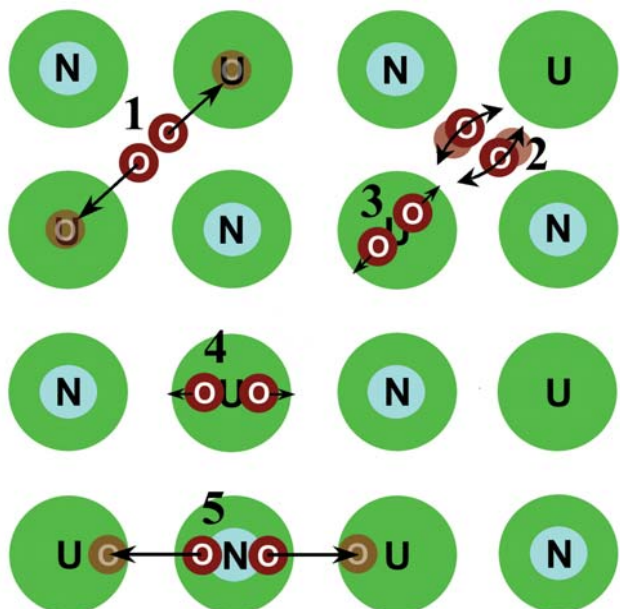

Fig. 1. (Color online). Schematic view of five different horizontal configurations for the O$_2$ molecule adsorption on UN surface: 1) atop the hollow site oriented towards the nearest surface U ions, 2) atop the hollow site oriented towards the nearest surface N ions, 3) atop the surface U ions oriented towards the next-nearest surface U ions, 4) atop the surface U ions oriented towards the nearest surface N ions, 5) atop the surface N ions oriented towards the nearest surface U ions. We show that molecule spontaneous dissociation can occur when O$_2$ is located either atop the hollow site (1) or atop ion N (5).

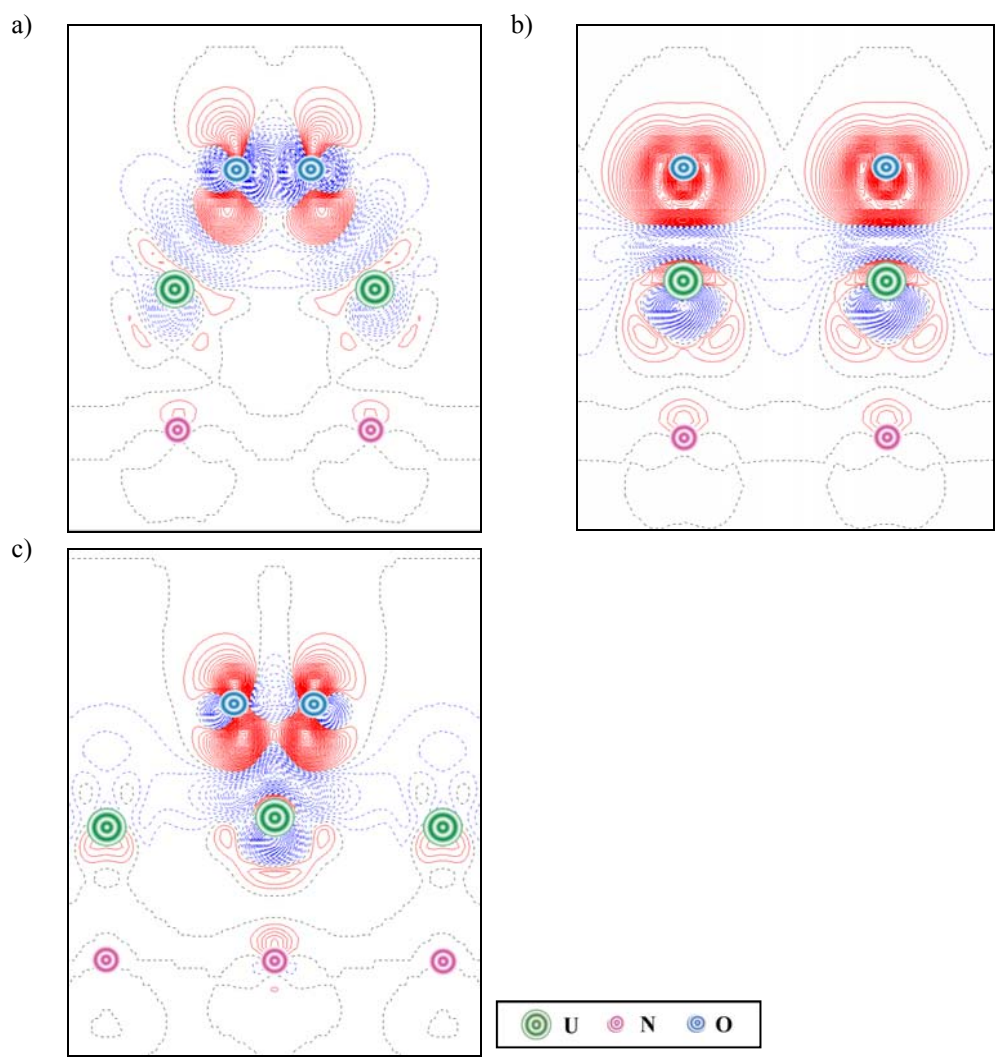

Fig. 2. (Color online). The difference electron density maps $\Delta\rho(\mathbf{r})$ (the total density of the interface minus the sum of densities of substrate and adsorbate with optimized interfacial geometry) for (a) the $O_2$ molecule upon the hollow position oriented to the nearest surface U ions, (b) after its dissociation in the configuration 1 (Fig. 1) with O atoms atop the surface U ions and (c) for the $O_2$ molecule atop the surface U ion in the configuration 3 (Fig. 1). Solid (red) and dashed (blue) isolines correspond to positive (excess) and negative (deficiency) electron density, respectively. Isodensity increment is 0.003 $e$ Å$^{-3}$.

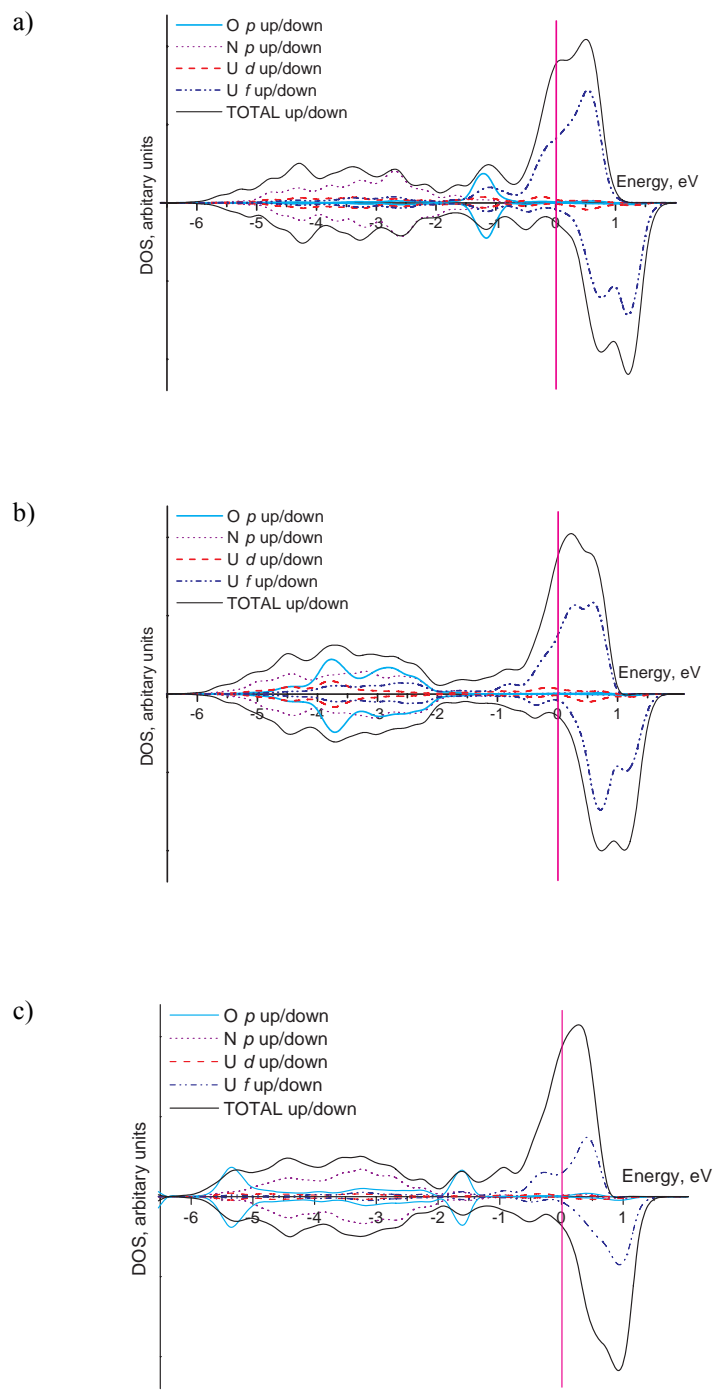

Fig. 3. (Color online). The total and projected densities of states for three configurations of $O_2$ molecule as in Fig. 2 (the same a, b, c). The orbital projections of both O atoms as well as the nearest N and U ions are shown. The highest peaks have been normalized to the same value, whereas a convolution of individual energy levels has been plotted using the Gaussian functions with a half-width of 0.2 eV.

Table 1. The calculated binding ($E_{bind}$, Eq. (1)) and dissociation $E_{diss}$ energies (eV), geometry ($z$, $\Delta z$) and charges ($q$) for configurations of molecular and spontaneous dissociative chemisorption of oxygen molecule aboth the UN(001) substrate. Numbers in brackets correspond to the configurations shown in Fig. 1.The calculated binding energy for a free $O_2$ molecule in the triplet state is 6.06 eV and a bond length of 1.31 Å (*cf.* with experimental values of 5.12 eV and 1.21 Å [24] respectively).

| Position | | $E_{bind}$ per O atom, eV | $z^a$, Å | $E_{diss}$, eV | $q(O)$, $e$ | $q(U1^b)$, $e$ | $q(U2^c)$, $e$ | $q(N^d)$, $e$ | $\Delta z^e(U1)$, Å | $\Delta z^e(U2)$, Å | $\Delta z^e(N)$, Å |
|---|---|---|---|---|---|---|---|---|---|---|---|
| hollow (1) | molecular adsorption | 3.03 | 1.893 | - | -0.465 | 1.913 | 1.762 | -1.533 | -0.0496 | -0.0496 | 0.02498 |
| | after dissociation | 6.04 | 1.957 | 3.01 | -0.978 | 2.053 | 1.978 | -1.577 | 0.075 | 0.068 | -0.133 |
| atop U | towards next-nearest U (3) | 4.00 | 2.18 | - | -0.5905 | 2.042 | 1.836 | -1.6065 | 0.176 | -0.048 | -0.096 |
| | towards nearest N (4) | 4.18 | 2.14 | - | -0.578 | 2.0485 | 1.827 | -1.6248 | 0.123 | -0.051 | -0.106 |
| atop N (5) | molecular adsorption | 2.67 | 2.020 | - | -0.5685 | 1.8675 | 1.8322 | -1.3537 | -0.0496 | -0.0496 | 0.025 |
| | after dissociation | 5.85 | 1.955 | 3.18 | -0.979 | 2.115 | 1.876 | -1.580 | 0.073 | 0.021 | -0.201 |

$^a$ $z$ is the height of O atoms respectively the non-relaxed UN substrate,

$^b$ U1 the nearest surface U ion,

$^c$ U2 the next-nearest surface U ion,

$^d$ N the nearest surface N ion,

$^e$ $\Delta z$ the additional vertical shifts of the same surface ions from their positions in the absence of adsorbed oxygen.